\pdfoutput=1

\documentclass[sigconf]{acmart}

\AtBeginDocument{%
  \providecommand\BibTeX{{%
    \normalfont B\kern-0.5em{\scshape i\kern-0.25em b}\kern-0.8em\TeX}}}

\setcopyright{acmcopyright}
\copyrightyear{2018}
\acmYear{2018}
\acmDOI{10.1145/1122445.1122456}

\acmConference[Web Intelligence '19]{IEEE/WIC/ACM International Conference on Web Intelligence}{October 14--17, 2019}{Thessaloniki, Greece}
\acmPrice{}
\acmISBN{}

\begin{document}

\title{Blockchain based access control systems: State of the art and challenges}

%
\author{Sara Rouhani}
\email{sara.rouhani@usask.ca}
\authornotemark[1]
\affiliation{%
  \department {Department of Computer Science}
  \institution{University of Saskatchewan}
  \city{Saskatoon}
  \state{SK}
  \country{Canada}
  }

\author{Ralph Deters}
\email{deters@cs.usask.ca}
\affiliation{%
  \department {Department of Computer Science}
  \institution{University of Saskatchewan}
 \city{Saskatoon}
  \state{SK}
  \country{Canada}
}

\renewcommand{\shortauthors}{Trovato and Tobin, et al.}

\begin{abstract}
Access control is a mechanism in computer security that regulates access to the system resources. The current access control systems face many problems, such as the presence of the third-party, inefficiency, and lack of privacy. These problems can be addressed by
blockchain, the technology that received major attention in recent years and has many potentials. In this study, we
overview the problems of the current access control systems, and then, we explain how blockchain can help to solve them. We also present an
overview of access control studies and proposed platforms in the different domains.
This paper presents the state of the art and the challenges of blockchain-based access control systems. 

\end{abstract}


\keywords{blockchain, access control, distributed, privacy}


\maketitle

\section{Introduction}
Blockchain applications initially were limited to the cryptocurrencies and financial transactions. Invention of smart contracts leads to development of more divers applications \cite{cai2018decentralized}, such as healthcare \cite{azaria2016medrec, dagher2018ancile, xia2017medshare, RouhaniMedichain}, IoT \cite{samaniego2016blockchain, pinno2017controlchain, ouaddah2016fairaccess, dukkipati2018decentralized, deters2017decentralized, samaniego2017internet}, supply chain \cite{chen2017blockchain, bocek2017blockchains, korpela2017digital}. In our previous study \cite{SmartContractReview}, after reviewing many research studies based on blockchain and smart contracts, we noticed that the primary focus of many presented applications is providing an efficient and secure access control mechanism.

Access control is a required security part of almost all applications. Blockchain specific characteristics such as immutability, durability, auditability, and reliability lead to considering blockchain as a supplementary solution for access control systems. 

Access control systems are applied to regulate access to the system's resources and it is the fundamental part of computer security. Access control is usually enforced against a set of authorization based on system policies. 

In this study we aim to provide the answer for following questions.
\begin{itemize}
\item What are the problems with current access control systems?
\item How blockchain can help to solve these problems?
\item What are the challenges for implementing an access control system based on blockchain?
\item What are the gaps in the related studies?  
\end{itemize}

In Section two, we investigate current access control systems problems and explain how blockchain can address them. We overview the related research studies and categorize them based on different domains and applied access control method in section three. In section four, we discuss the challenges of implementing an access control system using blockchain. Finally, in section five, we present the summary of the paper. We aim to provide a comprehensive picture with the details of architecture, implementation, and the challenges. 

\section{Traditional access control system problems and blockchain key benefits} 
In this section, we discuss the problems of current access control systems and how we can address them with blockchain. 

Jemel et al. \cite{jemel2017decentralized} mention a couple of problems in centralized access control systems.  As there is a third party, which has access to the data, the risk of privacy leakage exists. Also, a central party is in charge to control the access, so the risk of single point of failure also exists. This study presents an access control mechanism with a temporal dimension to solve these problems and adapts a blockchain-based solution for verifying access permissions. 
 
Attribute-based Encryption method \cite{sahai2005fuzzy} also has some problems such as privacy leakage from the private key generator (PKG) \cite{hur2011attribute} and single point of failure as mentioned before.  Wang et al. \cite{wang2018blockchain} introduce a framework for data sharing and access control to address this problem by implementing decentralized storage. 

Current solutions for managing access control in multi administrative domains are not efficient. Based on Paillisse et al. \cite{Distributed2019} static approaches are not scalable and granular and PKI-based systems are difficult to manage. They suggest distributing and recording access policies in a permissioned blockchain. Conifer \cite{dong2018conifer} is also another PKI system based on blockchain to achieve security without trusted third parties. 

In cloud federation also sharing data between multiple organization is a concern from users privacy perspective \cite{alansari2017distributed}. Keeping the personal data related to the users' identities private, while giving them access to the shared data, is the main concern. Alansari et al. propose an attribute-based access control system based on symmetric key encryption. The system checks users attributes with access control policies to grant access permissions to the data belong to the federated organization, while it keeps the users attribute private from the federated organization. This study suggests blockchain and trusted execution environment to preserve the integrity of the policy evaluation process. 

The users of mobile applications always concern about privacy issues as they usually must give access to their private information. Enigma is an access control management system based on Ethereum blockchain which aims to solve this problem \cite{zyskind2015decentralizing}. The presented framework addresses three main concerns: data ownership, data transparency \& auditability, and fine-grained access control. The system is designed in a way that users are able to control their own personal data and make the process of access to their data transparent. Also, the users can modify or revoke access permissions to their personal data without uninstalling the mobile application. Figure 1 shows the overview of  Enigma framework. The system also contains three distributed databases: a blockchain, a Distributed Hash Table (DHT), and a Multi-Party Computation (MPC), which fragments data into smaller meaningless chunks and distribute it between nodes without replication.
\begin{figure}[!b]
  \centering
  \includegraphics[width=0.8\linewidth]{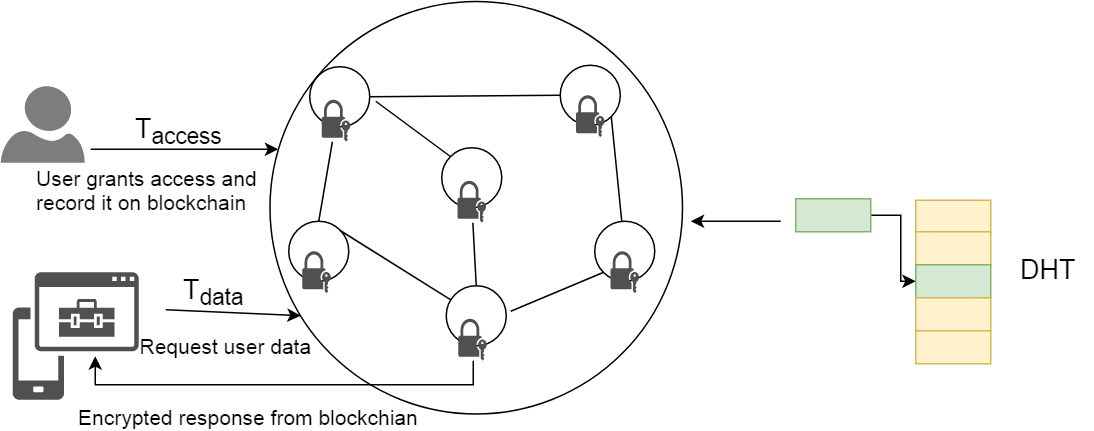}
  \caption{Enigma framework  \cite{zyskind2015decentralizing}}
\end{figure}

Privacy is not only a problem for the users of the mobile applications, in many access control systems the privacy is not guaranteed when the users grant access to their personal information in order to obtain access to the specific service or share resources. Table one shows research studies that aim to solve the privacy problem in an access control system using blockchain technology. 
 
\begin{table}[!t]
\small
\caption{Privacy based research studies}
\begin{tabular}{p{3.7cm}|p{4cm}}

Paper                                                                                                                                                                                                                          & Privacy Context                                      \\ \hline
Zyskind et al. \cite{zyskind2015decentralizing}                                                                                                                                                               & Mobile applications.                                 \\ \hline
Alansari et al. \cite{alansari2017privacy, alansari2017distributed}                                                                                                                                           & Cloud federation                                     \\ \hline
\begin{tabular}[c]{@{}l@{}}Ouaddah et al. \cite{ouaddah2017towards}, \\Pinno et al. \cite{ pinno2017controlchain},\\ Dukkipati \cite{dukkipati2018decentralized}, \\ Ma et al. \cite{ma2019privacy}\end{tabular} & \begin{tabular}[c]{@{}l@{}}Personal data generated by \\ IoT devices\end{tabular}             \\ \hline
\begin{tabular}[c]{@{}l@{}}Dagher et al. \cite{dagher2018ancile},\\ Azaria et al. \cite{azaria2016medrec}, \\ Xia et al. \cite{xia2017medshare}\end{tabular}                  & Healthcare patients data                             \\ \hline
Le and Mutka \cite{le2018capchain}                                                                                                                                                                            & Pervasive environment                                \\ \hline
Xu et al. \cite{xu2017enabling}                                                                                                                                                                               & \begin{tabular}[c]{@{}l@{}}sharing economy applications \\ using public blockchain\end{tabular} \\ \hline
Yao et al. \cite{yao2019pbcert}                                                                                                                                                                               & Certificate validation                               \\ \hline
Es-Samaali and Outchakoucht \cite{EsSamaali2017ABA}                                                                                                                                                           & Big Data
\\ 
                                            
\end{tabular}
\end{table}

\section{blockchain-based access control system}
Blockchain has desirable features that make it a trustable alternative infrastructure for access control systems. The distributed nature of blockchain solves the problem of single point of failure and other centralized management problems. Also, by eliminating third parties, we do not need to be concern about privacy leakage from their side. In addition, we can access to a trustable and unmodifiable history logs. Consensus mechanisms are applied, so only valid transactions are recorded on blockchain. Furthermore, by using smart contracts, we can monitor and enforce access permissions under complex conditions. All of these features have motivated researchers to consider blockchain as an infrastructure for access control systems.

\begin{table*}
\small
\caption{A summary of blockchain-based access control applications}
\begin{tabular}{c|c|c|c}
Research paper        & Domain                           & Access control method                                                                              & Blockchain platform \\ \hline
Maesa et al. \cite{maesa2017blockchain} & General access control           & Attribute-based                                                                                    & Bitcoin             \\ \hline
Maesa et al. \cite{di2018blockchain} & General access control & attribute-based & Ethereum \\ \hline
Jemel and Serhrouchni \cite{jemel2017decentralized} & Data sharing                     & Attribute-based Encryption                                                                                    & MultiChain          \\ \hline
Wang et al. \cite{wang2018blockchain} & Data sharing                     & Attribute-based Encryption                                                                                    & Ethereum            \\ \hline
Zhu et al. \cite{zhu2018tbac}	& Resource sharing	& attribute-based	&	Bitcoin	\\ \hline
Hu et al. \cite{hu2018reputation}       & Knowledge sharing                & Fine-grained                                                                                       & -                   \\ \hline
Zhu et al .    \cite{zhu2018digital}       & Digital asset management         & Attribute-based                                                                                    & Bitcoin             \\ \hline
Ferdous et al. \cite{ferdous2017decentralised}       & Cloud federation                 & -                                                                                                  & Hyperledger Fabric  \\ \hline
Alansari et al.   \cite{alansari2017distributed}     & Cloud federation                 & Attribute-base                                                                                     & -                   \\ \hline
Zhang and Posland  \cite{zhang2018blockchain}   & Electronic Medical Record (EMR)  & Granular attribute-based                                                                           & -                   \\ \hline
Rouhani et al. \cite{RouhaniMedichain}   & Medical data sharing (MediCHain) & Role-based & Hyperledger Fabric  \\ \hline
Asaph et al.  \cite{azaria2016medrec}        & Medical data sharing (MedRec)    & Fine-grined                                                                                        & Ethereum            \\ \hline
Xia et al.   \cite{xia2017medshare}      & Medical data sharing (MedShare)  & -                                                                                                  & Bitcoin             \\ \hline
Dagher et al. \cite{dagher2018ancile}		& Medical data sharing (Ancile)	&	Role-based	&	Ethereum \\ \hline
Novo       \cite{novo2018blockchain}           & IoT                              & -                                                                                                  & Private Ethereum    \\ \hline
Deters \cite{deters2017decentralized}	& IoT	& -	& MultiChain \\ \hline
Dukkipati et al. \cite{dukkipati2018decentralized}		& IoT	& Attribute-based & -	\\ \hline
Pinno et al. \cite{pinno2017controlchain}	& IoT (ControlChain)	& attribute-based	& -	\\ \hline
Ouaddah et al.   \cite{ouaddah2016fairaccess}     & IoT (FairAccess)                 & \begin{tabular}[c]{@{}c@{}}Generic \end{tabular} & Bitcoin             \\ \hline
Rouhani et al.   \cite{rouhani2017physical}     & Physical access control          & Role-based                                                                                         & Hyperledger Fabric  \\ \hline
Es-Samaali    \cite{EsSamaali2017ABA}        & Big data management              & Attribute-based                                                                                    & Bitcoin             \\ \hline
Stanciu     \cite{stanciu2017blockchain}          & Edge computing                   & -                                                                                                  & Hyperledger Fabric  \\ \hline
Paillisse et al\cite {Distributed2019} & Multi-administrative domain & - & Hyperledger Fabric \\  \hline
Maesa et al. \cite{maesa2019blockchain} & General access control & Attribute-based & Ethereum \\ \hline
Ding et al. \cite{ding2019novel} & IoT & Attribute-based & Hyperledger Fabric \\ \hline
Ma et al. \cite{ma2019privacy} & IoT & General access control & Multiblockchain \\ \hline
Samaniego et al. \cite{samaniego2019access} & Plant Phenotyping data & General access control & Ethereum \\

\end{tabular}
\end{table*}

This section overviews blockchain-based access control studies and decentralized applications. Table 2 shows the these studies classified in different domains, their access control method and their applied blockchain platform.
\subsection{Blockchain-based access control from transactions to smart contracts}
Maesa et al. \cite{maesa2017blockchain} initially represented a system by extending Bitcoin, which users can transparently observe access control policies on resources. This study uses attribute-based access control mechanism and eXtensible Access Control Markup Language (XACML) to define policies \cite{godik2002oasis} and store arbitrary data on Bitcoin. They used OP-RETURN script opcode and MULTISIG transactions \cite{shirriff2014hidden}. In their next study, they considered smart contracts to enforce access control policies instead of simple transactions \cite{di2018blockchain}. They have implemented a proof of concept using XACML policies and Ethereum platform. In their recent study \cite {maesa2019blockchain}, they have added more details and completed their previous works by explaining how the components of an access control system can be adopted in blockchain infrastructure. Also, in order to evaluate the feasibility and performance of the represented system, they have defined a scenario where smart contracts are considered as resources that need to be protected and access to them is restricted. By employing smart contracts, they were able to add more flexibility, details, and efficiency to their implemented system.

\subsection{Data sharing access control}
Jemel and Serhrouchni \cite {jemel2017decentralized} suggest using blockchain as an infrastructure for shared data access control management system. A proof of concept has been implemented using Multichain platform and CP-ABE (Ciphertext-Policy Attribute-Based Encryption) access control schema \cite{bethencourt2007ciphertext}. The analysis result indicates that timely CP-ABE performs better in terms of performance in comparison with timely access control list. As we expected, timely CP-ABE implementation without blockchain is more efficient than blockchain-based solutions, but using blockchain provides security and privacy benefits such as auditing, non-repudiation, as well as no single point of failure.  

Wang et al. \cite {wang2018blockchain} introduce a framework for data sharing and access control. The framework includes IPFS decentralized storage system, Ethereum blockchain, and Attribute-Based Encryption (ABE). The only one who has access to the secret key is the data owner. Ethereum blockchain has been applied for managing the private keys. There are two main smart contracts: data sharing contract that is deployed by the data owner and includes methods to register a user who need access to the specific data belong to the owner of the contract and dataUser contract that is deployed by data requester to invoke the search function defined in data sharing contract to view the search results.

Similarly, Hu et al. \cite{hu2018reputation} propose a Reputation Based Knowledge Sharing system to protect the copyright using fine-grained access control. The system includes three main roles:  Questioner, Answerer, and Bystander. The Questioner is the one who designs a question. The answerer is one who is an expert to answer the question and receives rewards from Bystander. The Bystander is the one who is willing to pay a small fee in order to get access to the shared knowledge. 

\subsection{Access control for cloud federation}
Ferdous et al. present a Decentralised Runtime Access Monitoring System (DRAMS) to guarantee the reliability of the access control component in cloud federations dynamically \cite{ferdous2017decentralised}. The represented architecture comprises three components: Logger, Smart contract, and Analyser. Logger component includes "Probing agents" records and forwards data to generate access logs and "Logging interface (LI)". Smart contracts capture logs and carry out monitoring by comparing logs to create dynamic access permissions. Analyzer investigates access permissions based on the system policies. 

Similarly, Alansari et al. \cite {alansari2017distributed} present an identity and access control management framework for cloud federation while keeps the attributes related to the users' identity private by using the OCBE protocol \cite{li2006oacerts}. Although the federation party, which owns the data, do not have access to the users' attributes, the user can access the requested data if the user has access to the data based on the system policy. The paper suggests the combination of blockchain and Intel SGX (Trusted Execution Environment) \cite{xing2016intel} for maintaining the integrity of the system. The users' identity attributes and the system access control policies are managed through smart contract and stored on blockchain. The encrypted data should be stored off-chain and in order to preserve the integrity, the cryptographic policy protocol runs in the trusted environment. 

\subsection{Access control across multiple organizations}
Cruz et al. \cite{cruz2018rbac} have designed a platform for role based access control to utilize across multiple organization using Ethereum blockchain and Solidity smart contracts. It has implemented a smart contract to initialize the roles and the challenge-response protocol to authenticate the ownership of roles and user verification.
The smart contract includes the following functions:

addUsert(u.EOA, u.role, u.notes) and removeUser(u.EOA) to assign a role to or revoke a role from the specific user identified by EOA (Externally owned Account or public key in Ethereum). addEndorser(eu.EOA, eu.notes) and removeEndorser(eu.EOA) to add and remove endorser and function changeStatus() to change the status of deactivated smart contracts.

Challenge-response protocol is utilized for the authentication of the users, who request a service from another organization based on her/his role. This protocol has five steps: declaration, information check, challenge response, and response verification. In summary, a user requests a service corresponding to his/her own roles from another organization. After initial information check, the organization sends an arbitrary data and ask the user to sign it and user responses with the signature. Finally, the authentication confirms after receiving valid signature.

\subsection{Access control for shared blockchains}
ChainAchor \cite{hardjono2016verifiable} is a blockchain platform that enforces access control for users who submit transactions. This paper introduces ChainAchor as a platform to solve the problem of identity and access control in the shared permissioned blockchains. Shared permissioned blockchain is a permissioned blockchain that is shared between multiple distinct organizations. Identity privacy, access control, and optional disclosure \& transaction privacy are challenging issues in shared permissioned blockchains. ChainAnchor consensus method looks for the public-key of the sender of the transaction in a database include all the identities information and it forces access control based on that. The identities of the users are anonymous completely and cannot be disclosed by anyone in the system. 
\subsection{Access control and self-Sovereign identities}
Users of digital identity systems suffer from lacks of privacy. When they request a service, for proving their identities, all the metadata attached to their digital identities become accessible to service provider. In self-sovereign identity system, the owners of digital identities are able to control the data related to their digital identities and their personal data attach to them using blockchain. Yan et al. \cite{yan2017bc} present a hierarchical secret sharing scheme for general access structure in blockchain to achieve self-sovereign digital identity metadata sharing. Finally, the paper introduces blockchain as a ledger for openPDS (Open Personal Data Store) \cite{de2014openpds}. 
\section{Challenge Discussion}
This section discusses the existing challenges and possible solutions.

Off-chain and on-chain integration:
Blockchain is not a suitable structure for storing a big volume of data, so, the data must store in secure off-chain storage and the access policies, the hash of the data, and references to the data record on blockchain. The secure integration between on-chain and off-chain is challenging. Using trusted hardware technology such as intel SGX can be considered to keep the integrity of the system. 

Blockchain vulnerability: 
Besides all the attractive advantages of blockchain, it is still difficult to implement non-vulnerable smart contracts \cite{luu2016making}. Subsequently, designing methods and tools to improve the security of smart contracts and blockchain is one of the most competitive fields in blockchain \cite{bhargavan2016formal, amani2018towards, bragagnolo2018smartinspect}. 

Transaction transparency:
One of the main reasons that blockchain became popular was providing transactional transparency; however, this is not desirable from the enterprise perspective and privacy point of view. That is why we have observed the advent of permissioned platforms, which support transaction privacy and private data. This sacrifices pure decentralization and adopt hybrid centralized and decentralized solutions. 

Performance:
Blockchain stores all the recorded transactions and data on all peers. The performance of execution and validation of transactions have been improved recently by introducing lighter consensus mechanism and more efficient transaction processing flow in blockchain platforms such as Hyperledger Fabric. Despite recent studies in improving the performance of blockchain \cite{dickerson2017adding, anjana2018efficient} still, the performance of the blockchain-based solutions cannot compete with the current centralized solutions. As we can see most of the studies compare the performance of their presented system with other blockchain-based platforms, not current decentralized solutions. In order to solve the performance and scalability problem, Ma et al. \cite{ma2019privacy} suggest an architecture based on multiple blockchains on cloud environment. The represented architecture consists of multiple layers, including device layer, edge networking layer, fog layer, core network layer, and cloud layer.  Simulations results indicate that balancing the load of block mining into multiple layers improves the performance of the system by reducing the transaction collection time and block mining time.

Access control methods:
Current access control methods which are static might be inadequate for future systems \cite{Calo2018} and more dynamic access control methods, one in which resources define their own access control, might be required. Integrating blockchain with dynamic access control approaches could be an interesting area to investigate in the future.  

\section{summary and future works}
In this paper, we explained the required concepts related to blockchain, smart contracts, platforms, and access control methods. The problem of the current access control systems and how blockchain features can solve these problems have been discussed.

Represented blockchain-based access control architectures and systems have been classified based on domain, access control method, and blockchain platforms. These systems have been tailored based on system requirements. These studies \cite{maesa2017blockchain, wang2018blockchain, RouhaniMedichain, azaria2016medrec} have focused on designing a user-centric system, which owners of the data can define and enforce access control policies directly. \cite{ferdous2017decentralised, maesa2019blockchain} systems have focused on auditability characteristic and trusted logging provided by blockchain to design a reliable access control system. From the transactional perspective, \cite{maesa2017blockchain, jemel2017decentralized, zhu2018digital} use only transactions to store access control attributes on blockchain, while \cite{EsSamaali2017ABA, di2018blockchain, maesa2019blockchain, xia2017medshare, RouhaniMedichain, ouaddah2017towards, wang2018blockchain, hu2018reputation, ferdous2017decentralised, alansari2017distributed,cruz2018rbac} applied smart contracts to exploit its advantages such as flexibility and automatically enforcing access control policies. Also, the challenges and future directions have been discussed in this paper. 

In our future work, we plan to present an access control service oriented \cite{liu2008management} architecture and implementation based on blockchain.

\bibliographystyle{ACM-Reference-Format}
\bibliography{sample-sigconf}


\begin{thebibliography}{60}


\ifx \showCODEN    \undefined \def \showCODEN     #1{\unskip}     \fi
\ifx \showDOI      \undefined \def \showDOI       #1{#1}\fi
\ifx \showISBNx    \undefined \def \showISBNx     #1{\unskip}     \fi
\ifx \showISBNxiii \undefined \def \showISBNxiii  #1{\unskip}     \fi
\ifx \showISSN     \undefined \def \showISSN      #1{\unskip}     \fi
\ifx \showLCCN     \undefined \def \showLCCN      #1{\unskip}     \fi
\ifx \shownote     \undefined \def \shownote      #1{#1}          \fi
\ifx \showarticletitle \undefined \def \showarticletitle #1{#1}   \fi
\ifx \showURL      \undefined \def \showURL       {\relax}        \fi
\providecommand\bibfield[2]{#2}
\providecommand\bibinfo[2]{#2}
\providecommand\natexlab[1]{#1}
\providecommand\showeprint[2][]{arXiv:#2}

\bibitem[\protect\citeauthoryear{Alansari, Paci, Margheri, and
  Sassone}{Alansari et~al\mbox{.}}{2017b}]%
        {alansari2017privacy}
\bibfield{author}{\bibinfo{person}{Shorouq Alansari}, \bibinfo{person}{Federica
  Paci}, \bibinfo{person}{Andrea Margheri}, {and} \bibinfo{person}{Vladimiro
  Sassone}.} \bibinfo{year}{2017}\natexlab{b}.
\newblock \showarticletitle{Privacy-preserving access control in cloud
  federations}. In \bibinfo{booktitle}{\emph{Cloud Computing (CLOUD), 2017 IEEE
  10th International Conference on}}. IEEE, \bibinfo{pages}{757--760}.
\newblock


\bibitem[\protect\citeauthoryear{Alansari, Paci, and Sassone}{Alansari
  et~al\mbox{.}}{2017a}]%
        {alansari2017distributed}
\bibfield{author}{\bibinfo{person}{Shorouq Alansari}, \bibinfo{person}{Federica
  Paci}, {and} \bibinfo{person}{Vladimiro Sassone}.}
  \bibinfo{year}{2017}\natexlab{a}.
\newblock \showarticletitle{A distributed access control system for cloud
  federations}. In \bibinfo{booktitle}{\emph{Distributed Computing Systems
  (ICDCS), 2017 IEEE 37th International Conference on}}. IEEE,
  \bibinfo{pages}{2131--2136}.
\newblock


\bibitem[\protect\citeauthoryear{Amani, B{\'e}gel, Bortin, and Staples}{Amani
  et~al\mbox{.}}{2018}]%
        {amani2018towards}
\bibfield{author}{\bibinfo{person}{Sidney Amani}, \bibinfo{person}{Myriam
  B{\'e}gel}, \bibinfo{person}{Maksym Bortin}, {and} \bibinfo{person}{Mark
  Staples}.} \bibinfo{year}{2018}\natexlab{}.
\newblock \showarticletitle{Towards verifying ethereum smart contract bytecode
  in Isabelle/HOL}. In \bibinfo{booktitle}{\emph{Proceedings of the 7th ACM
  SIGPLAN International Conference on Certified Programs and Proofs}}. ACM,
  \bibinfo{pages}{66--77}.
\newblock


\bibitem[\protect\citeauthoryear{Anjana, Kumari, Peri, Rathor, and
  Somani}{Anjana et~al\mbox{.}}{2018}]%
        {anjana2018efficient}
\bibfield{author}{\bibinfo{person}{Parwat~Singh Anjana}, \bibinfo{person}{Sweta
  Kumari}, \bibinfo{person}{Sathya Peri}, \bibinfo{person}{Sachin Rathor},
  {and} \bibinfo{person}{Archit Somani}.} \bibinfo{year}{2018}\natexlab{}.
\newblock \showarticletitle{An Efficient Framework for Concurrent Execution of
  Smart Contracts}.
\newblock \bibinfo{journal}{\emph{arXiv preprint arXiv:1809.01326}}
  (\bibinfo{year}{2018}).
\newblock


\bibitem[\protect\citeauthoryear{Azaria, Ekblaw, Vieira, and Lippman}{Azaria
  et~al\mbox{.}}{2016}]%
        {azaria2016medrec}
\bibfield{author}{\bibinfo{person}{Asaph Azaria}, \bibinfo{person}{Ariel
  Ekblaw}, \bibinfo{person}{Thiago Vieira}, {and} \bibinfo{person}{Andrew
  Lippman}.} \bibinfo{year}{2016}\natexlab{}.
\newblock \showarticletitle{Medrec: Using blockchain for medical data access
  and permission management}. In \bibinfo{booktitle}{\emph{Open and Big Data
  (OBD), International Conference on}}. IEEE, \bibinfo{pages}{25--30}.
\newblock


\bibitem[\protect\citeauthoryear{Bethencourt, Sahai, and Waters}{Bethencourt
  et~al\mbox{.}}{2007}]%
        {bethencourt2007ciphertext}
\bibfield{author}{\bibinfo{person}{John Bethencourt}, \bibinfo{person}{Amit
  Sahai}, {and} \bibinfo{person}{Brent Waters}.}
  \bibinfo{year}{2007}\natexlab{}.
\newblock \showarticletitle{Ciphertext-policy attribute-based encryption}. In
  \bibinfo{booktitle}{\emph{Security and Privacy, 2007. SP'07. IEEE Symposium
  on}}. IEEE, \bibinfo{pages}{321--334}.
\newblock


\bibitem[\protect\citeauthoryear{Bhargavan, Delignat-Lavaud, Fournet,
  Gollamudi, Gonthier, Kobeissi, Kulatova, Rastogi, Sibut-Pinote, Swamy,
  et~al\mbox{.}}{Bhargavan et~al\mbox{.}}{2016}]%
        {bhargavan2016formal}
\bibfield{author}{\bibinfo{person}{Karthikeyan Bhargavan},
  \bibinfo{person}{Antoine Delignat-Lavaud}, \bibinfo{person}{C{\'e}dric
  Fournet}, \bibinfo{person}{Anitha Gollamudi}, \bibinfo{person}{Georges
  Gonthier}, \bibinfo{person}{Nadim Kobeissi}, \bibinfo{person}{Natalia
  Kulatova}, \bibinfo{person}{Aseem Rastogi}, \bibinfo{person}{Thomas
  Sibut-Pinote}, \bibinfo{person}{Nikhil Swamy}, {et~al\mbox{.}}}
  \bibinfo{year}{2016}\natexlab{}.
\newblock \showarticletitle{Formal verification of smart contracts: Short
  paper}. In \bibinfo{booktitle}{\emph{Proceedings of the 2016 ACM Workshop on
  Programming Languages and Analysis for Security}}. ACM,
  \bibinfo{pages}{91--96}.
\newblock


\bibitem[\protect\citeauthoryear{Bocek, Rodrigues, Strasser, and Stiller}{Bocek
  et~al\mbox{.}}{2017}]%
        {bocek2017blockchains}
\bibfield{author}{\bibinfo{person}{Thomas Bocek}, \bibinfo{person}{Bruno~B
  Rodrigues}, \bibinfo{person}{Tim Strasser}, {and} \bibinfo{person}{Burkhard
  Stiller}.} \bibinfo{year}{2017}\natexlab{}.
\newblock \showarticletitle{Blockchains everywhere-a use-case of blockchains in
  the pharma supply-chain}. In \bibinfo{booktitle}{\emph{2017 IFIP/IEEE
  Symposium on Integrated Network and Service Management (IM)}}. IEEE,
  \bibinfo{pages}{772--777}.
\newblock


\bibitem[\protect\citeauthoryear{Bragagnolo, Rocha, Denker, and
  Ducasse}{Bragagnolo et~al\mbox{.}}{2018}]%
        {bragagnolo2018smartinspect}
\bibfield{author}{\bibinfo{person}{Santiago Bragagnolo},
  \bibinfo{person}{Henrique Rocha}, \bibinfo{person}{Marcus Denker}, {and}
  \bibinfo{person}{St{\'e}phane Ducasse}.} \bibinfo{year}{2018}\natexlab{}.
\newblock \showarticletitle{SmartInspect: solidity smart contract inspector}.
  In \bibinfo{booktitle}{\emph{2018 International Workshop on Blockchain
  Oriented Software Engineering (IWBOSE)}}. IEEE, \bibinfo{pages}{9--18}.
\newblock


\bibitem[\protect\citeauthoryear{Cai, Wang, Ernst, Hong, Feng, and Leung}{Cai
  et~al\mbox{.}}{2018}]%
        {cai2018decentralized}
\bibfield{author}{\bibinfo{person}{Wei Cai}, \bibinfo{person}{Zehua Wang},
  \bibinfo{person}{Jason~B Ernst}, \bibinfo{person}{Zhen Hong},
  \bibinfo{person}{Chen Feng}, {and} \bibinfo{person}{Victor~CM Leung}.}
  \bibinfo{year}{2018}\natexlab{}.
\newblock \showarticletitle{Decentralized applications: The
  blockchain-empowered software system}.
\newblock \bibinfo{journal}{\emph{IEEE Access}}  \bibinfo{volume}{6}
  (\bibinfo{year}{2018}), \bibinfo{pages}{53019--53033}.
\newblock


\bibitem[\protect\citeauthoryear{Calo, Verma, Chakraborty, Bertino, Lupu, and
  Cirincione}{Calo et~al\mbox{.}}{2018}]%
        {Calo2018}
\bibfield{author}{\bibinfo{person}{Seraphin Calo}, \bibinfo{person}{Dinesh
  Verma}, \bibinfo{person}{Supriyo Chakraborty}, \bibinfo{person}{Elisa
  Bertino}, \bibinfo{person}{Emil Lupu}, {and} \bibinfo{person}{Gregory
  Cirincione}.} \bibinfo{year}{2018}\natexlab{}.
\newblock \showarticletitle{Self-Generation of Access Control Policies}.
\newblock  (\bibinfo{year}{2018}), \bibinfo{pages}{39--47}.
\newblock


\bibitem[\protect\citeauthoryear{Chen, Shi, Ren, Yan, Shi, and Zhang}{Chen
  et~al\mbox{.}}{2017}]%
        {chen2017blockchain}
\bibfield{author}{\bibinfo{person}{Si Chen}, \bibinfo{person}{Rui Shi},
  \bibinfo{person}{Zhuangyu Ren}, \bibinfo{person}{Jiaqi Yan},
  \bibinfo{person}{Yani Shi}, {and} \bibinfo{person}{Jinyu Zhang}.}
  \bibinfo{year}{2017}\natexlab{}.
\newblock \showarticletitle{A blockchain-based supply chain quality management
  framework}. In \bibinfo{booktitle}{\emph{2017 IEEE 14th International
  Conference on e-Business Engineering (ICEBE)}}. IEEE,
  \bibinfo{pages}{172--176}.
\newblock


\bibitem[\protect\citeauthoryear{Cruz, Kaji, and Yanai}{Cruz
  et~al\mbox{.}}{2018}]%
        {cruz2018rbac}
\bibfield{author}{\bibinfo{person}{Jason~Paul Cruz}, \bibinfo{person}{Yuichi
  Kaji}, {and} \bibinfo{person}{Naoto Yanai}.} \bibinfo{year}{2018}\natexlab{}.
\newblock \showarticletitle{RBAC-SC: Role-Based Access Control Using Smart
  Contract}.
\newblock \bibinfo{journal}{\emph{IEEE Access}}  \bibinfo{volume}{6}
  (\bibinfo{year}{2018}), \bibinfo{pages}{12240--12251}.
\newblock


\bibitem[\protect\citeauthoryear{Dagher, Mohler, Milojkovic, and
  Marella}{Dagher et~al\mbox{.}}{2018}]%
        {dagher2018ancile}
\bibfield{author}{\bibinfo{person}{Gaby~G Dagher}, \bibinfo{person}{Jordan
  Mohler}, \bibinfo{person}{Matea Milojkovic}, {and}
  \bibinfo{person}{Praneeth~Babu Marella}.} \bibinfo{year}{2018}\natexlab{}.
\newblock \showarticletitle{Ancile: Privacy-preserving framework for access
  control and interoperability of electronic health records using blockchain
  technology}.
\newblock \bibinfo{journal}{\emph{Sustainable Cities and Society}}
  \bibinfo{volume}{39} (\bibinfo{year}{2018}), \bibinfo{pages}{283--297}.
\newblock


\bibitem[\protect\citeauthoryear{De~Montjoye, Shmueli, Wang, and
  Pentland}{De~Montjoye et~al\mbox{.}}{2014}]%
        {de2014openpds}
\bibfield{author}{\bibinfo{person}{Yves-Alexandre De~Montjoye},
  \bibinfo{person}{Erez Shmueli}, \bibinfo{person}{Samuel~S Wang}, {and}
  \bibinfo{person}{Alex~Sandy Pentland}.} \bibinfo{year}{2014}\natexlab{}.
\newblock \showarticletitle{openpds: Protecting the privacy of metadata through
  safeanswers}.
\newblock \bibinfo{journal}{\emph{PloS one}} \bibinfo{volume}{9},
  \bibinfo{number}{7} (\bibinfo{year}{2014}), \bibinfo{pages}{e98790}.
\newblock


\bibitem[\protect\citeauthoryear{Deters}{Deters}{2017}]%
        {deters2017decentralized}
\bibfield{author}{\bibinfo{person}{Ralph Deters}.}
  \bibinfo{year}{2017}\natexlab{}.
\newblock \showarticletitle{Decentralized Access Control with Distributed
  Ledgers}.
\newblock \bibinfo{journal}{\emph{University of Saskatchewan Cloud Robotics}}
  (\bibinfo{year}{2017}).
\newblock


\bibitem[\protect\citeauthoryear{DI~FRANCESCO~MAESA, Mori, and
  Ricci}{DI~FRANCESCO~MAESA et~al\mbox{.}}{2018}]%
        {di2018blockchain}
\bibfield{author}{\bibinfo{person}{Damiano DI~FRANCESCO~MAESA},
  \bibinfo{person}{Paolo Mori}, {and} \bibinfo{person}{LAURA~EMILIA Ricci}.}
  \bibinfo{year}{2018}\natexlab{}.
\newblock \showarticletitle{Blockchain based access control services}. In
  \bibinfo{booktitle}{\emph{IEEE Symposium on Recent Advances on Blockchain and
  its Applications, Canada,}}.
\newblock


\bibitem[\protect\citeauthoryear{Dickerson, Gazzillo, Herlihy, and
  Koskinen}{Dickerson et~al\mbox{.}}{2017}]%
        {dickerson2017adding}
\bibfield{author}{\bibinfo{person}{Thomas Dickerson}, \bibinfo{person}{Paul
  Gazzillo}, \bibinfo{person}{Maurice Herlihy}, {and} \bibinfo{person}{Eric
  Koskinen}.} \bibinfo{year}{2017}\natexlab{}.
\newblock \showarticletitle{Adding concurrency to smart contracts}. In
  \bibinfo{booktitle}{\emph{Proceedings of the ACM Symposium on Principles of
  Distributed Computing}}. ACM, \bibinfo{pages}{303--312}.
\newblock


\bibitem[\protect\citeauthoryear{Ding, Cao, Li, Fan, and Li}{Ding
  et~al\mbox{.}}{2019}]%
        {ding2019novel}
\bibfield{author}{\bibinfo{person}{Sheng Ding}, \bibinfo{person}{Jin Cao},
  \bibinfo{person}{Chen Li}, \bibinfo{person}{Kai Fan}, {and}
  \bibinfo{person}{Hui Li}.} \bibinfo{year}{2019}\natexlab{}.
\newblock \showarticletitle{A Novel Attribute-Based Access Control Scheme Using
  Blockchain for IoT}.
\newblock \bibinfo{journal}{\emph{IEEE Access}}  \bibinfo{volume}{7}
  (\bibinfo{year}{2019}), \bibinfo{pages}{38431--38441}.
\newblock


\bibitem[\protect\citeauthoryear{Dong, Kim, and Boutaba}{Dong
  et~al\mbox{.}}{2018}]%
        {dong2018conifer}
\bibfield{author}{\bibinfo{person}{Yuhao Dong}, \bibinfo{person}{Woojung Kim},
  {and} \bibinfo{person}{Raouf Boutaba}.} \bibinfo{year}{2018}\natexlab{}.
\newblock \showarticletitle{Conifer: centrally-managed PKI with
  blockchain-rooted trust}. In \bibinfo{booktitle}{\emph{IEEE International
  Conference on Blockchain (Blockchain)}}.
\newblock


\bibitem[\protect\citeauthoryear{Dukkipati, Zhang, and Cheng}{Dukkipati
  et~al\mbox{.}}{2018}]%
        {dukkipati2018decentralized}
\bibfield{author}{\bibinfo{person}{Chethana Dukkipati},
  \bibinfo{person}{Yunpeng Zhang}, {and} \bibinfo{person}{Liang~Chieh Cheng}.}
  \bibinfo{year}{2018}\natexlab{}.
\newblock \showarticletitle{Decentralized, BlockChain Based Access Control
  Framework for the Heterogeneous Internet of Things}. In
  \bibinfo{booktitle}{\emph{Proceedings of the Third ACM Workshop on
  Attribute-Based Access Control}}. ACM, \bibinfo{pages}{61--69}.
\newblock


\bibitem[\protect\citeauthoryear{Es-Samaali, Outchakoucht, and
  Leroy}{Es-Samaali et~al\mbox{.}}{2017}]%
        {EsSamaali2017ABA}
\bibfield{author}{\bibinfo{person}{Hamza Es-Samaali}, \bibinfo{person}{Aissam
  Outchakoucht}, {and} \bibinfo{person}{J.~P. Leroy}.}
  \bibinfo{year}{2017}\natexlab{}.
\newblock \showarticletitle{A Blockchain-based Access Control for Big Data}.
\newblock


\bibitem[\protect\citeauthoryear{Ferdous, Margheri, Paci, Yang, and
  Sassone}{Ferdous et~al\mbox{.}}{2017}]%
        {ferdous2017decentralised}
\bibfield{author}{\bibinfo{person}{Md~Sadek Ferdous}, \bibinfo{person}{Andrea
  Margheri}, \bibinfo{person}{Federica Paci}, \bibinfo{person}{Mu Yang}, {and}
  \bibinfo{person}{Vladimiro Sassone}.} \bibinfo{year}{2017}\natexlab{}.
\newblock \showarticletitle{Decentralised runtime monitoring for access control
  systems in cloud federations}. In \bibinfo{booktitle}{\emph{Distributed
  Computing Systems (ICDCS), 2017 IEEE 37th International Conference on}}.
  IEEE, \bibinfo{pages}{2632--2633}.
\newblock


\bibitem[\protect\citeauthoryear{Godik and Moses}{Godik and Moses}{2002}]%
        {godik2002oasis}
\bibfield{author}{\bibinfo{person}{Simon Godik} {and} \bibinfo{person}{Tim
  Moses}.} \bibinfo{year}{2002}\natexlab{}.
\newblock \showarticletitle{Oasis extensible access control markup language
  (xacml)}.
\newblock \bibinfo{journal}{\emph{OASIS Committee Secification
  cs-xacml-specification-1.0}} (\bibinfo{year}{2002}).
\newblock


\bibitem[\protect\citeauthoryear{Hardjono and Pentland}{Hardjono and
  Pentland}{2016}]%
        {hardjono2016verifiable}
\bibfield{author}{\bibinfo{person}{Thomas Hardjono} {and}
  \bibinfo{person}{Alex~Sandy Pentland}.} \bibinfo{year}{2016}\natexlab{}.
\newblock \showarticletitle{Verifiable Anonymous Identities and Access Control
  in Permissioned Blockchains}.
\newblock \bibinfo{journal}{\emph{manuscript in preparation}}
  (\bibinfo{year}{2016}).
\newblock


\bibitem[\protect\citeauthoryear{Hu, Hou, Chen, Weng, and Li}{Hu
  et~al\mbox{.}}{2018}]%
        {hu2018reputation}
\bibfield{author}{\bibinfo{person}{Shuang Hu}, \bibinfo{person}{Lin Hou},
  \bibinfo{person}{Gongliang Chen}, \bibinfo{person}{Jian Weng}, {and}
  \bibinfo{person}{Jianhua Li}.} \bibinfo{year}{2018}\natexlab{}.
\newblock \showarticletitle{Reputation-based Distributed Knowledge Sharing
  System in Blockchain}. In \bibinfo{booktitle}{\emph{Proceedings of the 15th
  EAI International Conference on Mobile and Ubiquitous Systems: Computing,
  Networking and Services}}. ACM, \bibinfo{pages}{476--481}.
\newblock


\bibitem[\protect\citeauthoryear{Hur and Noh}{Hur and Noh}{2011}]%
        {hur2011attribute}
\bibfield{author}{\bibinfo{person}{Junbeom Hur} {and} \bibinfo{person}{Dong~Kun
  Noh}.} \bibinfo{year}{2011}\natexlab{}.
\newblock \showarticletitle{Attribute-based access control with efficient
  revocation in data outsourcing systems}.
\newblock \bibinfo{journal}{\emph{IEEE Transactions on Parallel and Distributed
  Systems}} \bibinfo{volume}{22}, \bibinfo{number}{7} (\bibinfo{year}{2011}),
  \bibinfo{pages}{1214--1221}.
\newblock


\bibitem[\protect\citeauthoryear{Jemel and Serhrouchni}{Jemel and
  Serhrouchni}{2017}]%
        {jemel2017decentralized}
\bibfield{author}{\bibinfo{person}{Mayssa Jemel} {and} \bibinfo{person}{Ahmed
  Serhrouchni}.} \bibinfo{year}{2017}\natexlab{}.
\newblock \showarticletitle{Decentralized access control mechanism with
  temporal dimension based on blockchain}. In \bibinfo{booktitle}{\emph{2017
  IEEE 14th International Conference on e-Business Engineering (ICEBE)}}. IEEE,
  \bibinfo{pages}{177--182}.
\newblock


\bibitem[\protect\citeauthoryear{Korpela, Hallikas, and Dahlberg}{Korpela
  et~al\mbox{.}}{2017}]%
        {korpela2017digital}
\bibfield{author}{\bibinfo{person}{Kari Korpela}, \bibinfo{person}{Jukka
  Hallikas}, {and} \bibinfo{person}{Tomi Dahlberg}.}
  \bibinfo{year}{2017}\natexlab{}.
\newblock \showarticletitle{Digital supply chain transformation toward
  blockchain integration}. In \bibinfo{booktitle}{\emph{proceedings of the 50th
  Hawaii international conference on system sciences}}.
\newblock


\bibitem[\protect\citeauthoryear{Le and Mutka}{Le and Mutka}{2018}]%
        {le2018capchain}
\bibfield{author}{\bibinfo{person}{Tam Le} {and} \bibinfo{person}{Matt~W
  Mutka}.} \bibinfo{year}{2018}\natexlab{}.
\newblock \showarticletitle{CapChain: A Privacy Preserving Access Control
  Framework Based on Blockchain for Pervasive Environments}. In
  \bibinfo{booktitle}{\emph{2018 IEEE International Conference on Smart
  Computing (SMARTCOMP)}}. IEEE, \bibinfo{pages}{57--64}.
\newblock


\bibitem[\protect\citeauthoryear{Li and Li}{Li and Li}{2006}]%
        {li2006oacerts}
\bibfield{author}{\bibinfo{person}{Jiangtao Li} {and} \bibinfo{person}{Ninghui
  Li}.} \bibinfo{year}{2006}\natexlab{}.
\newblock \showarticletitle{OACerts: Oblivious attribute certificates}.
\newblock \bibinfo{journal}{\emph{IEEE Transactions on Dependable and Secure
  Computing}} \bibinfo{volume}{3}, \bibinfo{number}{4} (\bibinfo{year}{2006}),
  \bibinfo{pages}{340--352}.
\newblock


\bibitem[\protect\citeauthoryear{Liu and Deters}{Liu and Deters}{2008}]%
        {liu2008management}
\bibfield{author}{\bibinfo{person}{Dong Liu} {and} \bibinfo{person}{Ralph
  Deters}.} \bibinfo{year}{2008}\natexlab{}.
\newblock \showarticletitle{Management of service-oriented systems}.
\newblock \bibinfo{journal}{\emph{Service Oriented Computing and Applications}}
  \bibinfo{volume}{2}, \bibinfo{number}{2-3} (\bibinfo{year}{2008}),
  \bibinfo{pages}{51--64}.
\newblock


\bibitem[\protect\citeauthoryear{Luu, Chu, Olickel, Saxena, and Hobor}{Luu
  et~al\mbox{.}}{2016}]%
        {luu2016making}
\bibfield{author}{\bibinfo{person}{Loi Luu}, \bibinfo{person}{Duc-Hiep Chu},
  \bibinfo{person}{Hrishi Olickel}, \bibinfo{person}{Prateek Saxena}, {and}
  \bibinfo{person}{Aquinas Hobor}.} \bibinfo{year}{2016}\natexlab{}.
\newblock \showarticletitle{Making smart contracts smarter}. In
  \bibinfo{booktitle}{\emph{Proceedings of the 2016 ACM SIGSAC Conference on
  Computer and Communications Security}}. ACM, \bibinfo{pages}{254--269}.
\newblock


\bibitem[\protect\citeauthoryear{Ma, Shi, and Li}{Ma et~al\mbox{.}}{2019}]%
        {ma2019privacy}
\bibfield{author}{\bibinfo{person}{Mingxin Ma}, \bibinfo{person}{Guozhen Shi},
  {and} \bibinfo{person}{Fenghua Li}.} \bibinfo{year}{2019}\natexlab{}.
\newblock \showarticletitle{Privacy-Oriented Blockchain-Based Distributed Key
  Management Architecture for Hierarchical Access Control in the IoT Scenario}.
\newblock \bibinfo{journal}{\emph{IEEE Access}}  \bibinfo{volume}{7}
  (\bibinfo{year}{2019}), \bibinfo{pages}{34045--34059}.
\newblock


\bibitem[\protect\citeauthoryear{Maesa, Mori, and Ricci}{Maesa
  et~al\mbox{.}}{2017}]%
        {maesa2017blockchain}
\bibfield{author}{\bibinfo{person}{Damiano Di~Francesco Maesa},
  \bibinfo{person}{Paolo Mori}, {and} \bibinfo{person}{Laura Ricci}.}
  \bibinfo{year}{2017}\natexlab{}.
\newblock \showarticletitle{Blockchain based access control}. In
  \bibinfo{booktitle}{\emph{IFIP International Conference on Distributed
  Applications and Interoperable Systems}}. Springer,
  \bibinfo{pages}{206--220}.
\newblock


\bibitem[\protect\citeauthoryear{Maesa, Mori, and Ricci}{Maesa
  et~al\mbox{.}}{2019}]%
        {maesa2019blockchain}
\bibfield{author}{\bibinfo{person}{Damiano Di~Francesco Maesa},
  \bibinfo{person}{Paolo Mori}, {and} \bibinfo{person}{Laura Ricci}.}
  \bibinfo{year}{2019}\natexlab{}.
\newblock \showarticletitle{A blockchain based approach for the definition of
  auditable Access Control systems}.
\newblock \bibinfo{journal}{\emph{Computers \& Security}}
  (\bibinfo{year}{2019}).
\newblock


\bibitem[\protect\citeauthoryear{Novo}{Novo}{2018}]%
        {novo2018blockchain}
\bibfield{author}{\bibinfo{person}{Oscar Novo}.}
  \bibinfo{year}{2018}\natexlab{}.
\newblock \showarticletitle{Blockchain meets IoT: An architecture for scalable
  access management in IoT}.
\newblock \bibinfo{journal}{\emph{IEEE Internet of Things Journal}}
  \bibinfo{volume}{5}, \bibinfo{number}{2} (\bibinfo{year}{2018}),
  \bibinfo{pages}{1184--1195}.
\newblock


\bibitem[\protect\citeauthoryear{Ouaddah, Abou~Elkalam, and
  Ait~Ouahman}{Ouaddah et~al\mbox{.}}{2016}]%
        {ouaddah2016fairaccess}
\bibfield{author}{\bibinfo{person}{Aafaf Ouaddah}, \bibinfo{person}{Anas
  Abou~Elkalam}, {and} \bibinfo{person}{Abdellah Ait~Ouahman}.}
  \bibinfo{year}{2016}\natexlab{}.
\newblock \showarticletitle{FairAccess: a new Blockchain-based access control
  framework for the Internet of Things}.
\newblock \bibinfo{journal}{\emph{Security and Communication Networks}}
  \bibinfo{volume}{9}, \bibinfo{number}{18} (\bibinfo{year}{2016}),
  \bibinfo{pages}{5943--5964}.
\newblock


\bibitem[\protect\citeauthoryear{Ouaddah, Elkalam, and Ouahman}{Ouaddah
  et~al\mbox{.}}{2017}]%
        {ouaddah2017towards}
\bibfield{author}{\bibinfo{person}{Aafaf Ouaddah}, \bibinfo{person}{Anas~Abou
  Elkalam}, {and} \bibinfo{person}{Abdellah~Ait Ouahman}.}
  \bibinfo{year}{2017}\natexlab{}.
\newblock \showarticletitle{Towards a novel privacy-preserving access control
  model based on blockchain technology in IoT}.
\newblock In \bibinfo{booktitle}{\emph{Europe and MENA Cooperation Advances in
  Information and Communication Technologies}}. \bibinfo{publisher}{Springer},
  \bibinfo{pages}{523--533}.
\newblock


\bibitem[\protect\citeauthoryear{Paillisse, Subira, Lopez, Rodriguez-Natal,
  Ermagan, Maino, and Cabellos}{Paillisse et~al\mbox{.}}{2019}]%
        {Distributed2019}
\bibfield{author}{\bibinfo{person}{Jordi Paillisse}, \bibinfo{person}{jordi
  Subira}, \bibinfo{person}{Alber Lopez}, \bibinfo{person}{Alberto
  Rodriguez-Natal}, \bibinfo{person}{Vina Ermagan}, \bibinfo{person}{Fabio
  Maino}, {and} \bibinfo{person}{Albert Cabellos}.}
  \bibinfo{year}{2019}\natexlab{}.
\newblock \showarticletitle{Distributed Access Control with Blockchain}.
\newblock \bibinfo{journal}{\emph{arXiv preprint arXiv:1901.03568}}
  (\bibinfo{year}{2019}).
\newblock


\bibitem[\protect\citeauthoryear{Pinno, Gregio, and De~Bona}{Pinno
  et~al\mbox{.}}{2017}]%
        {pinno2017controlchain}
\bibfield{author}{\bibinfo{person}{Otto Julio~Ahlert Pinno},
  \bibinfo{person}{Andre Ricardo~Abed Gregio}, {and} \bibinfo{person}{Luis~CE
  De~Bona}.} \bibinfo{year}{2017}\natexlab{}.
\newblock \showarticletitle{ControlChain: Blockchain as a Central Enabler for
  Access Control Authorizations in the IoT}. In
  \bibinfo{booktitle}{\emph{GLOBECOM 2017-2017 IEEE Global Communications
  Conference}}. IEEE, \bibinfo{pages}{1--6}.
\newblock


\bibitem[\protect\citeauthoryear{Rouhani, Butterworth, Dimmond, Humphery, and
  Deters}{Rouhani et~al\mbox{.}}{2018a}]%
        {RouhaniMedichain}
\bibfield{author}{\bibinfo{person}{S. Rouhani}, \bibinfo{person}{L.
  Butterworth}, \bibinfo{person}{A.~D. Dimmond}, \bibinfo{person}{D.~G.
  Humphery}, {and} \bibinfo{person}{R. Deters}.}
  \bibinfo{year}{2018}\natexlab{a}.
\newblock \showarticletitle{MediChainTM: A Secure Decentralized Medical Data
  Asset Management System}. In \bibinfo{booktitle}{\emph{2018 IEEE
  International Conference on Internet of Things (iThings) and IEEE Green
  Computing and Communications (GreenCom) and IEEE Cyber, Physical and Social
  Computing (CPSCom) and IEEE Smart Data (SmartData)}}.
  \bibinfo{publisher}{IEEE}, \bibinfo{pages}{14757--14767}.
\newblock


\bibitem[\protect\citeauthoryear{Rouhani and Deters}{Rouhani and
  Deters}{2019}]%
        {SmartContractReview}
\bibfield{author}{\bibinfo{person}{Sara Rouhani} {and} \bibinfo{person}{Ralph
  Deters}.} \bibinfo{year}{2019}\natexlab{}.
\newblock \showarticletitle{Security, Performance, and Applications of Smart
  Contracts: A Systematic Survey}.
\newblock \bibinfo{journal}{\emph{IEEE ACCESS}}  \bibinfo{volume}{7}
  (\bibinfo{year}{2019}), \bibinfo{pages}{50759--50779}.
\newblock


\bibitem[\protect\citeauthoryear{Rouhani, pourheidari, and Deters}{Rouhani
  et~al\mbox{.}}{2018b}]%
        {rouhani2017physical}
\bibfield{author}{\bibinfo{person}{Sara Rouhani}, \bibinfo{person}{Vahid
  pourheidari}, {and} \bibinfo{person}{Ralph Deters}.}
  \bibinfo{year}{2018}\natexlab{b}.
\newblock \showarticletitle{Physical Access Control Management System Based on
  Permissioned Blockchain}. In \bibinfo{booktitle}{\emph{2018 IEEE
  International Conference on Internet of Things (iThings) and IEEE Green
  Computing and Communications (GreenCom) and IEEE Cyber, Physical and Social
  Computing (CPSCom) and IEEE Smart Data (SmartData)}}. IEEE,
  \bibinfo{pages}{1078--1083}.
\newblock


\bibitem[\protect\citeauthoryear{Sahai and Waters}{Sahai and Waters}{2005}]%
        {sahai2005fuzzy}
\bibfield{author}{\bibinfo{person}{Amit Sahai} {and} \bibinfo{person}{Brent
  Waters}.} \bibinfo{year}{2005}\natexlab{}.
\newblock \showarticletitle{Fuzzy identity-based encryption}. In
  \bibinfo{booktitle}{\emph{Annual International Conference on the Theory and
  Applications of Cryptographic Techniques}}. Springer,
  \bibinfo{pages}{457--473}.
\newblock


\bibitem[\protect\citeauthoryear{Samaniego and Deters}{Samaniego and
  Deters}{2016}]%
        {samaniego2016blockchain}
\bibfield{author}{\bibinfo{person}{Mayra Samaniego} {and}
  \bibinfo{person}{Ralph Deters}.} \bibinfo{year}{2016}\natexlab{}.
\newblock \showarticletitle{Blockchain as a Service for IoT}. In
  \bibinfo{booktitle}{\emph{2016 IEEE International Conference on Internet of
  Things (iThings) and IEEE Green Computing and Communications (GreenCom) and
  IEEE Cyber, Physical and Social Computing (CPSCom) and IEEE Smart Data
  (SmartData)}}. IEEE, \bibinfo{pages}{433--436}.
\newblock


\bibitem[\protect\citeauthoryear{Samaniego and Deters}{Samaniego and
  Deters}{2017}]%
        {samaniego2017internet}
\bibfield{author}{\bibinfo{person}{Mayra Samaniego} {and}
  \bibinfo{person}{Ralph Deters}.} \bibinfo{year}{2017}\natexlab{}.
\newblock \showarticletitle{Internet of smart things-iost: Using blockchain and
  clips to make things autonomous}. In \bibinfo{booktitle}{\emph{2017 IEEE
  international conference on cognitive computing (ICCC)}}. IEEE,
  \bibinfo{pages}{9--16}.
\newblock


\bibitem[\protect\citeauthoryear{Samaniego, Espana, and Deters}{Samaniego
  et~al\mbox{.}}{2019}]%
        {samaniego2019access}
\bibfield{author}{\bibinfo{person}{Mayra Samaniego}, \bibinfo{person}{Cristian
  Espana}, {and} \bibinfo{person}{Ralph Deters}.}
  \bibinfo{year}{2019}\natexlab{}.
\newblock \showarticletitle{Access Control Management for Plant Phenotyping
  Using Integrated Blockchain}. In \bibinfo{booktitle}{\emph{Proceedings of the
  2019 ACM International Symposium on Blockchain and Secure Critical
  Infrastructure}}. ACM, \bibinfo{pages}{39--46}.
\newblock


\bibitem[\protect\citeauthoryear{Shirriff}{Shirriff}{2014}]%
        {shirriff2014hidden}
\bibfield{author}{\bibinfo{person}{Ken Shirriff}.}
  \bibinfo{year}{2014}\natexlab{}.
\newblock \showarticletitle{Hidden surprises in the Bitcoin blockchain and how
  they are stored: Nelson Mandela, Wikileaks, photos, and Python software}.
\newblock \bibinfo{journal}{\emph{Ken Shirriff’s blog (accessed July 2017)
  http://www. righto. com/2014/02/ascii-bernanke-wikileaks-photographs. html}}
  (\bibinfo{year}{2014}).
\newblock


\bibitem[\protect\citeauthoryear{Stanciu}{Stanciu}{2017}]%
        {stanciu2017blockchain}
\bibfield{author}{\bibinfo{person}{Alexandru Stanciu}.}
  \bibinfo{year}{2017}\natexlab{}.
\newblock \showarticletitle{Blockchain based distributed control system for
  edge computing}. In \bibinfo{booktitle}{\emph{2017 21st International
  Conference on Control Systems and Computer Science (CSCS)}}. IEEE,
  \bibinfo{pages}{667--671}.
\newblock


\bibitem[\protect\citeauthoryear{Wang, Zhang, and Zhang}{Wang
  et~al\mbox{.}}{2018}]%
        {wang2018blockchain}
\bibfield{author}{\bibinfo{person}{Shangping Wang}, \bibinfo{person}{Yinglong
  Zhang}, {and} \bibinfo{person}{Yaling Zhang}.}
  \bibinfo{year}{2018}\natexlab{}.
\newblock \showarticletitle{A blockchain-based framework for data sharing with
  fine-grained access control in decentralized storage systems}.
\newblock \bibinfo{journal}{\emph{IEEE Access}}  \bibinfo{volume}{6}
  (\bibinfo{year}{2018}), \bibinfo{pages}{38437--38450}.
\newblock


\bibitem[\protect\citeauthoryear{Xia, Sifah, Asamoah, Gao, Du, and Guizani}{Xia
  et~al\mbox{.}}{2017}]%
        {xia2017medshare}
\bibfield{author}{\bibinfo{person}{Qi Xia}, \bibinfo{person}{Emmanuel~Boateng
  Sifah}, \bibinfo{person}{Kwame~Omono Asamoah}, \bibinfo{person}{Jianbin Gao},
  \bibinfo{person}{Xiaojiang Du}, {and} \bibinfo{person}{Mohsen Guizani}.}
  \bibinfo{year}{2017}\natexlab{}.
\newblock \showarticletitle{MeDShare: Trust-less medical data sharing among
  cloud service providers via blockchain}.
\newblock \bibinfo{journal}{\emph{IEEE Access}}  \bibinfo{volume}{5}
  (\bibinfo{year}{2017}), \bibinfo{pages}{14757--14767}.
\newblock


\bibitem[\protect\citeauthoryear{Xing, Shanahan, and Leslie-Hurd}{Xing
  et~al\mbox{.}}{2016}]%
        {xing2016intel}
\bibfield{author}{\bibinfo{person}{Bin~Cedric Xing}, \bibinfo{person}{Mark
  Shanahan}, {and} \bibinfo{person}{Rebekah Leslie-Hurd}.}
  \bibinfo{year}{2016}\natexlab{}.
\newblock \showarticletitle{Intel{\textregistered} Software Guard Extensions
  (Intel{\textregistered} SGX) Software Support for Dynamic Memory Allocation
  inside an Enclave}. In \bibinfo{booktitle}{\emph{Proceedings of the Hardware
  and Architectural Support for Security and Privacy 2016}}. ACM,
  \bibinfo{pages}{11}.
\newblock


\bibitem[\protect\citeauthoryear{Xu, Shah, Chen, Diallo, Gao, Lu, and Shi}{Xu
  et~al\mbox{.}}{2017}]%
        {xu2017enabling}
\bibfield{author}{\bibinfo{person}{Lei Xu}, \bibinfo{person}{Nolan Shah},
  \bibinfo{person}{Lin Chen}, \bibinfo{person}{Nour Diallo},
  \bibinfo{person}{Zhimin Gao}, \bibinfo{person}{Yang Lu}, {and}
  \bibinfo{person}{Weidong Shi}.} \bibinfo{year}{2017}\natexlab{}.
\newblock \showarticletitle{Enabling the sharing economy: Privacy respecting
  contract based on public blockchain}. In
  \bibinfo{booktitle}{\emph{Proceedings of the ACM Workshop on Blockchain,
  Cryptocurrencies and Contracts}}. ACM, \bibinfo{pages}{15--21}.
\newblock


\bibitem[\protect\citeauthoryear{Yan, Gan, and Riad}{Yan et~al\mbox{.}}{2017}]%
        {yan2017bc}
\bibfield{author}{\bibinfo{person}{Zhu Yan}, \bibinfo{person}{Guhua Gan}, {and}
  \bibinfo{person}{Khaled Riad}.} \bibinfo{year}{2017}\natexlab{}.
\newblock \showarticletitle{BC-PDS: protecting privacy and self-sovereignty
  through BlockChains for OpenPDS}. In \bibinfo{booktitle}{\emph{2017 IEEE
  Symposium on Service-Oriented System Engineering (SOSE)}}. IEEE,
  \bibinfo{pages}{138--144}.
\newblock


\bibitem[\protect\citeauthoryear{Yao, Chen, He, Du, Zhu, and Chen}{Yao
  et~al\mbox{.}}{2019}]%
        {yao2019pbcert}
\bibfield{author}{\bibinfo{person}{Shixiong Yao}, \bibinfo{person}{Jing Chen},
  \bibinfo{person}{Kun He}, \bibinfo{person}{Ruiying Du},
  \bibinfo{person}{Tianqing Zhu}, {and} \bibinfo{person}{Xin Chen}.}
  \bibinfo{year}{2019}\natexlab{}.
\newblock \showarticletitle{PBCert: Privacy-Preserving Blockchain-Based
  Certificate Status Validation Toward Mass Storage Management}.
\newblock \bibinfo{journal}{\emph{IEEE Access}}  \bibinfo{volume}{7}
  (\bibinfo{year}{2019}), \bibinfo{pages}{6117--6128}.
\newblock


\bibitem[\protect\citeauthoryear{Zhang and Poslad}{Zhang and Poslad}{2018}]%
        {zhang2018blockchain}
\bibfield{author}{\bibinfo{person}{Xiaoshuai Zhang} {and}
  \bibinfo{person}{Stefan Poslad}.} \bibinfo{year}{2018}\natexlab{}.
\newblock \showarticletitle{Blockchain Support for Flexible Queries with
  Granular Access Control to Electronic Medical Records (EMR)}. In
  \bibinfo{booktitle}{\emph{2018 IEEE International Conference on
  Communications (ICC)}}. IEEE, \bibinfo{pages}{1--6}.
\newblock


\bibitem[\protect\citeauthoryear{Zhu, Qin, Gan, Shuai, and Chu}{Zhu
  et~al\mbox{.}}{2018a}]%
        {zhu2018tbac}
\bibfield{author}{\bibinfo{person}{Yan Zhu}, \bibinfo{person}{Yao Qin},
  \bibinfo{person}{Guohua Gan}, \bibinfo{person}{Yang Shuai}, {and}
  \bibinfo{person}{William Cheng-Chung Chu}.} \bibinfo{year}{2018}\natexlab{a}.
\newblock \showarticletitle{TBAC: transaction-based access control on
  blockchain for resource sharing with cryptographically decentralized
  authorization}. In \bibinfo{booktitle}{\emph{2018 IEEE 42nd Annual Computer
  Software and Applications Conference (COMPSAC)}}, Vol.~\bibinfo{volume}{1}.
  IEEE, \bibinfo{pages}{535--544}.
\newblock


\bibitem[\protect\citeauthoryear{Zhu, Qin, Zhou, Song, Liu, and Chu}{Zhu
  et~al\mbox{.}}{2018b}]%
        {zhu2018digital}
\bibfield{author}{\bibinfo{person}{Yan Zhu}, \bibinfo{person}{Yao Qin},
  \bibinfo{person}{Zhiyuan Zhou}, \bibinfo{person}{Xiaoxu Song},
  \bibinfo{person}{Guowei Liu}, {and} \bibinfo{person}{William Cheng-Chung
  Chu}.} \bibinfo{year}{2018}\natexlab{b}.
\newblock \showarticletitle{Digital Asset Management with Distributed
  Permission over Blockchain and Attribute-Based Access Control}. In
  \bibinfo{booktitle}{\emph{2018 IEEE International Conference on Services
  Computing (SCC)}}. IEEE, \bibinfo{pages}{193--200}.
\newblock


\bibitem[\protect\citeauthoryear{Zyskind, Nathan, et~al\mbox{.}}{Zyskind
  et~al\mbox{.}}{2015}]%
        {zyskind2015decentralizing}
\bibfield{author}{\bibinfo{person}{Guy Zyskind}, \bibinfo{person}{Oz Nathan},
  {et~al\mbox{.}}} \bibinfo{year}{2015}\natexlab{}.
\newblock \showarticletitle{Decentralizing privacy: Using blockchain to protect
  personal data}. In \bibinfo{booktitle}{\emph{Security and Privacy Workshops
  (SPW), 2015 IEEE}}. IEEE, \bibinfo{pages}{180--184}.
\newblock


\end{thebibliography}

%
\appendix

\end{document}